\documentstyle[osa,aplop,preprint,epsfig]{revtex}

\begin{document}
\preprint{UDEM-GPP-TH-95031}
\vskip.5truecm
\title{Baby Skyrmion Strings}

\author{by T. Gisiger and M. B. Paranjape}

\address{Groupe de Physique des Particules, Laboratoire de physique
nucl\'eaire,
Universit\'e de Montr\'eal C.P. 6128, succ. centre-ville, Montr\'eal,
Qu\'ebec, Canada, H3C 3J7}

\maketitle

\begin{abstract}
We provide analytical and numerical evidence of the existence of classically
stable, string-like configurations in a 2+1 dimensional analog of the Skyrme
model. The model contains a conserved topological charge usually called  the
baryon number. Our strings are non-topological solitons which have a constant
baryon number per
unit length. The energy per length containing one baryon is, however, less than
the energy of an isolated baryon (radially symmetric ``baby Skyrmion") in a
region of the parameter space, which suggests a degree of stability for our
configurations. In a limiting case, our configuration saturates a
Bogomolnyi-type bound and is degenerate in energy per baryon with the baby
Skyrmion. In another limiting case, the energies are still degenerate but do
not
saturate the corresponding Bogomolnyi-type bound. Nonetheless, we expect the
string to be stable here. Both limiting cases are solvable analytically.
\end{abstract}
\vskip 1.0cm
The Skyrme model\cite{1} is a non-linear sigma model containing topological
solitons which describe the low-energy dynamics of mesons and baryons. Its 2+1
dimensional analog was studied in various contexts. Here the fields take values
in the two-sphere $S^2$, hence the dynamical variables correspond to maps from
$R^2$ to $S^2$. Imposing that
all configurations go to a constant at spatial infinity effectively
compactifies the spatial $R^2$ also into an $S^2$, and hence all maps are
characterized by a topological charge corresponding to the winding number of
maps from $S^2$ to $S^2$.  With the addition of the
Hopf term to the action, it was shown that the solitons have fractional spin
and
statistics\cite{2}. The stability and dynamics of the solitons of the
model, called baby Skyrmions, have been
studied\cite{3} and it was shown that stable localised solitons exist with the
incorporation of a Skyrme term and a mass term. For a study of many solutions
of
the $CP^1$ model in 2+1 dimensions, see reference [4]. Soliton-soliton
scattering has been studied in reference [5] which gives a simpler analog of
the
corresponding phenomenon in the usual Skyrme model\cite{6}. In a condensed
matter context it was shown that 2+1 dimension Skyrmions are also relevant in
quantum Hall systems and anyonic superconductors\cite{7}.

The energy functional is given by the usual kinetic term
of the $O(3)$ non-linear sigma model\cite{8}, a four derivative term which is
analogous to the Skyrme term\cite{9}$^,$\cite{3}, and a mass term (or
interaction with an external magnetic field) which actually breaks the
symmetry and picks out a vacuum\cite{10},
\begin{equation}
E = {1\over 2}\int d^2\vec x\Bigr[\partial_i \vec\phi(x)\cdot\partial_i
\vec\phi(\vec x) +
\bigl(\partial_1 \vec\phi(\vec x)\times\partial_2 \vec\phi(\vec x)\bigr)^2 +
\mu^2 \bigl(\vec n - \vec\phi(\vec x)\bigr)^2 \Bigl].\label{eq1}
\end{equation}
Here $\phi^a$, $a=1,2,3$ are the components of a unit vector $\vec\phi$,
$\vec\phi(\vec x)\cdot\vec\phi(\vec x)=1$,
and $\vec n$ is a constant unit vector taken for convenience to be $(0,0,1)$.
The kinetic term along with the Skyrme term are not sufficient to stabilize a
baby Skyrmion contrary to the usual Skyrme model.
The kinetic term in 2+1 dimensions enjoys (suffers
from) conformal invariance and the baby Skyrmion can always reduce its energy
by inflating (infinitely).  Hence one adds the mass term which limits the
size of the baby Skyrmion. The usual Skyrme term of course prohibits the
collapse of the soliton.

The configuration giving rise to a baby Skyrmion with topological charge $N$
is (see the first article of reference [2])
\begin{equation}
\vec\phi(r,\theta) = (\sin{f(r)} \cos{N \theta},\sin{f(r)} \sin{N \theta},
\cos{f(r)})\label{eq3}
\end{equation}
where $f(r)$ goes from $\pi$ at the origin to zero at $+\infty$, and
where the topological charge, which by analogy, we call the baryon number, is
given by
\begin{equation}
B = {1\over 4\pi}\int d^2\vec x\, B(\vec x) = {1\over 4\pi}\int d^2\vec x\,
\,\vec\phi\cdot\partial_1
\vec\phi(\vec x)\times\partial_2\vec\phi(\vec x).\label{eq4}
\end{equation}
This gives (for $N = 1$) the ordinary, non-linear differential equation for
$f(r)$\cite{3}:
\begin{equation}
\begin{array} {l}
\Biggl( r + {\sin^2{f(r)}\over r}\Biggr) f''(r) +
\Biggl( 1 - {\sin^2{f(r)}\over r^2}+
{f'(r)\sin{f(r)}\cos{f(r)}\over r}\Biggr) f'(r)
\\
\qquad\qquad\qquad\qquad\qquad\qquad\qquad\qquad\qquad\qquad- {\sin{f(r)}
\cos{f(r)}\over r} - r \mu^2 \sin{f(r)} = 0.
\end{array}\label{eq5}
\end{equation}
This equation can only be integrated numerically, except in some limiting
cases.
A solution with unit baryon number $B$,
has $f(0) = \pi$, descending with a finite slope at the origin and arriving to
zero (mod $2\pi$) at infinity with an exponential, two-dimentional Yukawa-like
fall-off governed by the mass term. The solution is singular in the sense that
it is not differentiable at the origin, however, the energy density is well
defined everywhere.

To construct the string-like configuration, we simply re-interpret the radial
coordinate $r$ and the angular coordinate $\theta$ as two Cartesian
coordinates:
$(r,\theta)\rightarrow(x,y)$. This has the effect of laying the configurations
which previously occured along rays, from the origin to $r=+\infty$, in a
linear progression along the $y$ axis with the value $(0,0,-1)$ at $x=0$ and
the value $(0,0,1)$ (the vacuum) at $x=+\infty$. This way we obtain one
baryon per length $2\pi$ in the $y$ direction for the right half plane. Along
the $y$ axis the configuration has the value $(0,0,-1)$ which is not in the
vacuum direction. We must extend
our configuration into the left half plane in a way that we also reach the
vacuum at $x=-\infty$. The symmetric fashion of achieving this is to glue on to
the line at $x=0$ a configuration which corresponds to creating the same
string-like
configuration as before, however, after having performed a rotation by $\pi$.
This rotation has the effect of reversing the directions of $x$ and $y$. Indeed
the configuration, where we have rescaled $y$,
\begin{equation}
\vec \phi(x,y) = ( \sin{f(x)}\cos{{\pi\,y\,\hbox{sign}(x)\over L}},
\sin{f(x)}\sin{{\pi\,y\,\hbox{sign}(x)\over L}},
\cos{f(x)})\label{eq6}
\end{equation}
with $f(0)=\pi$, $f(\pm\infty)=0\,\,\hbox{mod}\,\,2 \pi$, has unit baryon
number per
length $L$ and has a baryon number density which is independent of $y$:
\begin{equation}
B(x,y) = {\pi\,\hbox{sign}(x)\over L} \sin{f(x)} f'(x) .\label{eq7}
\end{equation}
The function $f(x)$ satisfies the differential equation
\begin{equation}
\Biggl( 1 + \biggl({\pi\over L}\biggr)^2 \sin^2 f(x)\Biggr)f''(x)  +
\biggl({\pi\over L}\biggr)^2 {\sin{2 f(x)}\over 2} \biggl( f'(x)^2 - 1\biggr) -
\mu^2 \sin{f(x)} = 0.\label{eq8}
\end{equation}
The solution is found in the right half plane on imposing the boundary
condition $f(0)=\pi$, $f(+\infty)=0$. The continuation to the left half plane
is
done by either reflecting $f(-|x|) = f(|x|)$ so that $f(-\infty)=0$ or
continuing smoothly the solution which then interpolates from $f(0)=\pi$ to
$f(-\infty)=2\pi$. Either interpolation gives the same energy and baryon number
density.  Equation (\ref{eq8}) can actually be integrated analytically giving
rise to the quadrature:
\begin{equation}
\int\sqrt {1 + \pi^2/L^2 \sin^2{f}\over
  2 \mu^2 ( 1-\cos{f}) + \pi^2/L^2 \sin^2{f}} df= - x.\label{eq9}
\end{equation}
Equation (\ref{eq9}) is not terribly useful to obtain $f(x)$, since we still
must invert the
function defined in terms of the integral on the left hand side,
however, we may use it to obtain the energy per baryon as the following
integral:
\begin{equation}
E_{string} = 8 L \int^1_0 dy \Biggl( 1 + {4 \pi^2\over L^2} ( 1 - y^2 )y^2
\Biggr)^{1/2} \Biggl( \mu^2 + {\pi^2\over L^2} \,\,y^2
\Biggr)^{1/2}.\label{eq10}
\end{equation}
This allows for a numerical calculation of the energy per baryon without
recourse to numerical resolution of any differential equation. Subsequently we
minimize with respect to $L$ to find the actual minimum energy string
configuration.  We find $E_{string} = 1.55356\times 4\pi$ with $L_{string} =
3.4542$ for the value of the mass
parameter $\mu^2 = 0.1$ as chosen in the second article of reference [3]
whereas for the baby Skyrmion $E_{Skyrmion} = 1.564\times 4\pi$. Comparing
$E_{string}$ to $E_{Skyrmion}$ (obtained
numerically using the shooting method) for general $\mu^2$ we find the
following
curve for $E_{string}-E_{Skyrmion}$ (see Figure 1).
This means that the string will be stable
against disintegration into individual baryons in the region $0 < \mu < 1$.
The upper limit, $\mu = 1$, is obtained numerically with an error of less than
$0.002\%$, but we lack an analytical understanding of this fact. The string
with
baryon number $N$ is obtained from (\ref{eq6}) by the straightforward
substitution $y\rightarrow N y$. It is evident from the structure of the ansatz
that a string with baryon number $N$ has length $N L_{string}$ and energy $N
E_{string}$.

Taking the limit $\mu\rightarrow 0$ while rescaling the coordinates
appropriately, removes the Skyrme and mass terms, and we obtain the
conformally invariant $O(3)$ non-linear sigma model with energy
\begin{equation}
E = {1\over 2} \int d^2\vec x\,\, \partial_i \vec\phi(\vec x)\cdot\partial_i
\vec\phi(\vec x).\label{eq11}
\end{equation}
As is well known, this can be written in a form making the Bogomolnyi
bound\cite{11} apparent
\begin{equation}
E = {1\over 2} \int d^2\Biggl[\vec x\, \biggl( \partial_1\vec\phi(\vec x) \pm
\vec\phi\times\partial_2\vec\phi(\vec x)\biggr)^2 \pm 2
B(\vec x)\Biggr].\label{eq12}
\end{equation}
Hence $E \ge 4 \pi |B|$, the Bogomolnyi bound is saturated when
\begin{equation}
\partial_1\vec\phi(\vec x)  \pm
\vec\phi(\vec x)\times\partial_2\vec\phi(\vec x) =0,\label{eq13}
\end{equation}
depending on the sign of the baryon number.  This equation has a solution for
the string configuration, the equation of motion being
\begin{equation}
f'(x) \mp {\pi\,\hbox{sign}(x)\over L} \sin f(x) =0\label{eq131}
\end{equation}
with solution
\begin{equation}
f(x) = 2 \arctan\biggl(
\alpha\,e^{\pm\pi|x|/L} \biggr)\label{eq132}
\end{equation}
where $\alpha$ is an arbitrary scale parameter. The solution (\ref{eq132})
shows
that each half string becomes infinitely wide achieving $\pi$ at
$x=+\infty$ for the $+$ sign, and at $x=-\infty$ for the $-$ sign.
Equation (\ref{eq13}) also has a baby Skyrmion type solution, satisfying the
equation
\begin{equation}
f'(r) = \pm {\sin f(r)\over r}\label{eq133}
\end{equation}
and given by
\begin{equation}
f(r) = 2 \arctan{\Bigl[(\alpha r)^{\pm1}\Bigr]}.\label{eq14}
\end{equation}
The $N$ baryon generalization also satisfies the Bogomolnyi bound.
Since the string saturates the Bogomolnyi bound per baryon, it is degenerate
per baryon
with a configuration of an isolated $N$ baryon solution that also saturates
this
bound. Hence the string configuration is classically stable in
this limit.
To elaborate this further we simply calculate the energy for a configuration
$\vec\phi = \vec\phi_0+\vec\delta\phi$, where $\vec\phi_0$ satisfies the
equation
(\ref{eq131}) or (\ref{eq133}), using the expression (\ref{eq12})
\begin{equation}
\begin{array} {l}
E(\phi) = E(\phi_0) + \int d^2\vec x\, \biggl[\partial_1\delta\vec\phi \pm
\bigl(\vec\phi_0\times\partial_2\delta\vec\phi+
\delta\vec\phi\times\partial_2\vec\phi_0 +
\delta\vec\phi\times\partial_2\delta\vec\phi\bigr)\Biggr]^2
\\
\qquad\quad= E(\phi_0) + \delta E.
\end{array}\label{eq15}
\end{equation}
The energy of the fluctuation $\delta E$ is clearly a positive, semi-definite
quantity.

The other limit, essentially $\mu\rightarrow +\infty$ also yields an
interesting
and analytically solvable system. Here, if we scale the coordinates
appropriately while taking the limit, we can dispense with the
kinetic term, leaving the energy functional
\begin{equation}
E = {1\over 2}\int d^2\vec x\Bigl[\bigl(\partial_1\vec\phi
\times\partial_2\vec\phi\bigr)^2 +
\mu^2 (\vec n - \vec\phi)^2\Bigr].\label{eq16}
\end{equation}
The equations of motion with the ansatz (\ref{eq3}) and (\ref{eq6}) are equally
well integrable.  Indeed for the radially symmetric situation we get
\begin{equation}
f''(r) {\sin f(r)\over r} + f'(r)^2 {\cos f(r)\over r} - f'(r) {\sin f(r)\over
r^2} - \mu^2 r = 0\label{eq17}
\end{equation}
which integrates to
\begin{equation}
- \cos f(r) = {\mu^2 r^4\over 8} - \mu r^2 + 1,\label{eq18}
\end{equation}
with the boundary conditions $f(0)=\pi$, and $f(r)\rightarrow 0$ as $r$ becomes
large.  The interesting feature of this solution is that the field achieves
$(0,0,1)$ exactly at a finite radius $R$, $f(R)=0$.  We fix $R$ by minimizing
the energy which amounts to the condition that the configuration
achieves $(0,0,1)$ smoothly and yields $R = 2/\sqrt{\mu}$. We glue on to the
outside for $r>R$ the vacuum. The energy turns out to be
\begin{equation}
E_{Skyrmion} = {4\over 3} \,4 \pi\mu.\label{eq18}
\end{equation}
For the $N$ baryon ansatz (\ref{eq3})), this energy is simply multiplied by
$N$.
For the ansatz (\ref{eq6}), we obtain the following equation of motion
\begin{equation}
f''(x) \Biggl({\pi\over L}\Biggr)^2 \sin^2 f(x) +
f'(x)^ 2 \Biggl({\pi\over L}\Biggr)^2 {\sin 2f(x)\over 2}
- \mu^2 \sin f(x) = 0\label{eq19}
\end{equation}
which also is trivially integrated,
\begin{equation}
- \cos f(x) = \Biggl({\mu L\over\pi}\Biggr)^2 {x^2\over 2}
- 2 \Biggl({\mu L\over\pi}\Biggr) x + 1
\label{eq20}
\end{equation}
exhibiting the same, intriguing behaviour that the
soliton cuts off at a finite half-width $W = 2 \pi/(\mu L)$. The energy per
baryon is also given by
\begin{equation}
E_{string} = {4\over 3} \, 4 \pi\mu.\label{eq21}
\end{equation}
Interestingly enough, this does not depend on $L$, the length per baryon, and
the energy per baryon of a thin string is equal to that of a wide
string. The area per baryon is also independent of the value of $L$
\begin{equation}
A_{string} = 2\times L\times W={4\pi\over \mu}.\label{eq22}
\end{equation}
Amazingly enough this corresponds exactly to the area occupied by the radially
symmetric baby Skyrmion, $A_{Skyrmion}=\pi R^2=4\pi/\mu$.  We have not
been able to unearth the reason for this
apparent equality of areas, and also for the apparent shape invariance of
the string itself.  There is perhaps an underlying
area preserving diffeomorphism invariance in the model resulting in this
``incompressibility".

This limiting case does in fact contain a Bogomolnyi type bound, which is not
saturated.  Indeed,
\begin{equation}
E = {1\over 2}\int d^2\vec x\, \Bigl[\bigl( \partial_1\vec\phi\times\partial_2
\vec\phi \pm\mu( \vec n - \vec\phi) \bigr)^2 \pm 2 \mu^2 B(\vec x)\Bigr]
\label{eq23}
\end{equation}
and the extra cross term in the perfect square involving $\vec n$ integrates to
zero since it is a total divergence. Hence the energy per baryon satisfies
\begin{equation}
E \ge 4 \pi \,\mu |B|.\label{eq23}
\end{equation}
In any case, since our strings per baryon number are degenerate with an
isolated
baryon, and we actually do not expect a lower energy configuration in this
sector, our configurations seem to be stable against disintegration into
single baryons.

The classical stability analysis for general $\mu$ is not straightforward.  For
$\mu\in [0,1]$ we have found
that the string is energetically stable against disintegration into single
baryons.  This is not valid for disintegration into configurations
with $B=2$.  Taking the spherical ansatz for this ``baby deuteron", $n=2$ in
(\ref{eq3}) and solving (numerically) the corresponding differential equation
for $f(r)$, yields a configuration that is quite tightly bound,
\begin{equation}
E_{deuteron}=2 E_{Skyrmion} -\Delta. \label{eq25}
\end{equation}
For example with $\mu^2 =0.1$ we get $\Delta = 0.191\times4\pi$, as found in
the
second article of reference [3]. Hence, if not
classically, we
expect a string to be able to quantum
mechanically tunnel into a configuration of isolated lumps with $B=2$.
Calculation of the tunelling rate requires a detailed understanding of the
potential on the space of configurations, which for the moment is lacking.
In the two limiting cases
studied above, the baby deuteron is degenerate with the string (taken in
lengths with baryon number two) and we expect that it will be stable even
quantum mechanically (although in the second case there may be some other bound
configuration for the deuteron).

As there is no topological reason for the stability of our strings, even less
so than for usual cosmic strings which do enjoy and utilise the topology in the
transverse direction for part of their stability, an infinite string will
reduce its energy per unit length (to zero) by diluting itself (infinitely)
along its length. This is not the relevant consideration. We expect our
configurations to be important when considering the low energy excitations of
lumps with large, but finite baryon number, $N$.  The topology of $R^2$
indicates that the minimum energy configuration will be a localised, lump of
energy, in a field configuration with most likely some large symmetry group.
The excitation of this system will always preserve the baryon number.  Hence if
a string configuration is excited, the relevant energy consideration must be
done at fixed baryon number.  This boils down to considering the energy per
baryon, as we have done above.  Excitation of this matter will involve the
string configurations, be they
only meta-stable.  This should correspond to a different low energy phase for
the matter described by this energy functional.

\acknowledgments
We thank R. Mackenzie and P. Winternitz for useful discussions, and also W.C.
Chen and L. Gagnon for help with the numerical work. This
work supported in part by NSERC of Canada and FCAR of Qu\'ebec.

\begin{figure}
\vspace{-2.cm}
\leavevmode
\epsfxsize=0pt\epsfbox{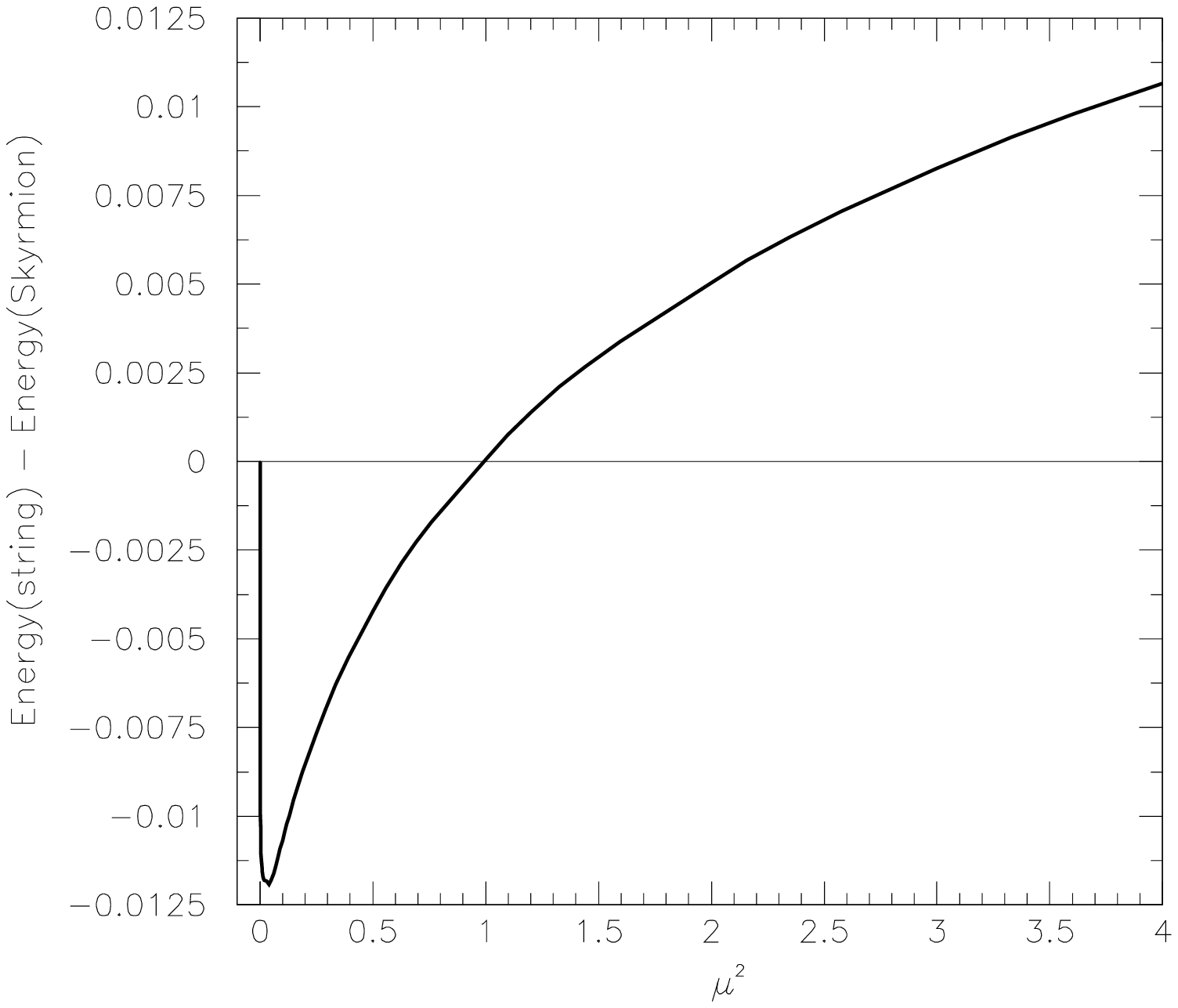}
\vspace{-2.cm}
\caption{$E_{string}-E_{Skyrmion}$ as a function of $\mu^2$.}
\end{figure}
\end{document}